\documentclass{llncs}
\usepackage{makeidx}  % allows for indexgeneration
\usepackage{color}
\usepackage{listings}
\usepackage{ulem}
\usepackage{aas_macros}
\usepackage{float}

\newfloat{query}{thp}{lop}
\floatname{query}{Query}

\newcommand\csharp{C$^\#$}
\newcommand\dotnet{.Net}

\definecolor{light-gray}{gray}{0.95}
\definecolor{light-blue}{cmyk}{0.1,0,0,0}
\begin{document}
\mainmatter              % start of the contributions
\title{SkyQuery: An Implementation of a Parallel Probabilistic Join Engine for Cross-Identification of Multiple Astronomical Databases}
\titlerunning{SkyQuery: A Parallel Probabilistic Join Engine}  % abbreviated title (for running head)
\author{L\'{a}szl\'{o} Dobos\inst{1,2} \and Tam\'{a}s Budav\'{a}ri\inst{2} \and Nolan Li\inst{2} \and \\
Alexander S. Szalay\inst{2} \and Istv\'{a}n Csabai\inst{1}}
\authorrunning{Dobos et al.} % abbreviated author list (for running head)
\tocauthor{L\'{a}szl\'{o} Dobos, Tam\'{a}s Budav\'{a}ri,
Alexander S. Szalay and Istv\'{a}n Csabai}
\institute{E\"{o}tv\"os Lor\'{a}nd University,
Department of Physics of Complex Systems, \\
H-1117 Budapest, Hungary,
\email{dobos@complex.elte.hu}
\and
The Johns Hopkins University,
Department of Physics \& Astronomy,\\
Baltimore, MD 21218, USA, \email{budavari@jhu.edu} 
}

\maketitle              % typeset the title of the contribution

\begin{abstract}
Multi-wavelength astronomical studies require cross-identi\-fication of detections of the same celestial objects in multiple catalogs based on spherical coordinates and other properties. Because of the large data volumes and spherical geometry, the symmetric N-way association of astronomical detections is a computationally intensive problem, even when sophisticated indexing schemes are used to exclude obviously false candidates. Legacy astronomical catalogs already contain detections of more than a hundred million objects while the ongoing and future surveys will produce catalogs of billions of objects with multiple detections of each at different times. The varying statistical error of position measurements, moving and extended objects, and other physical properties make it necessary to perform the cross-identification using a mathematically correct, proper Bayesian probabilistic algorithm, capable of including various priors. One time, pair-wise cross-identification of these large catalogs is not sufficient for many astronomical scenarios. Consequently, a novel system is necessary that can cross-identify multiple catalogs on-demand, efficiently and reliably. In this paper, we present our solution based on a cluster of commodity servers and ordinary relational databases. The cross-identification problems are formulated in a language based on SQL, but extended with special clauses. These special queries are partitioned spatially by coordinate ranges and compiled into a complex workflow of ordinary SQL queries. Workflows are then executed in a parallel framework using a cluster of servers hosting identical mirrors of the same data sets. The whole system consists of custom-written software modules for the probabilistic algorithms, cluster management, job queuing, high-performance bulk data copying, query parsing and optimization, and metadata management. The system is designed to be accessible via various types of web interfaces for human and program clients.
\end{abstract}

\section{Introduction} \label{sec:intro}

Increasingly large astronomical data warehouses are being built to support the needs of scientific collaborations. The International Virtual Observatory Alliance\footnote{\url{http://www.ivoa.net}} (IVOA) laid down the standards of transport data models and communication protocols to help the federation of geographically distributed data sets. The astronomer community is now working hard on the implementation of systems that will bring the power of petabyte-scale data warehouses and the information of hundreds of large archives to the desktops of scientists.

The key point in the federation of astronomical data sets is the cross-identification (cross-matching) of detections belonging to the same physical object. The detections are usually made by using different imaging filters or entirely different instruments. Due to the exponential growth in the data volume, our solution has to be scalable. Also, since the largest data sets will be geographically distributed and data co-location might not be an option in the future, any solution will have to be optimized for the networking. 

In this paper, we present a scalable solution for on-demand cross-matching of large catalogs hosted on a cluster of database servers. In the Sec.~\ref{sec:obs}, we explain the characteristic properties of astronomy catalogs. The cross-identification problem is introduced in Sec.~\ref{sec:xmatchintro}. Sec.~\ref{sec:sqlext} describes our SQL extensions to explicitly formulate the problem of coordinate based matching in a query request. The details of our hardware and software setup are highlighted in Sec.~\ref{sec:hwsw}. In Sec.~\ref{sec:implementation} we focus on the most important aspects of the implementation.  Sec.~\ref{sec:summary} concludes the paper, and outlines specific future work.

\section{Astronomical observations} \label{sec:obs}

Today's high-performance astronomical \textit{imaging instruments} (ground-based or space-borne telescopes with their custom-built cameras) are mostly operated in \textit{survey mode}, i.e. significantly large regions of the sky are mapped systematically. Every telescope is designed to work in a certain range of the electromagnetic spectrum, thus the division of fields of astronomy according to these regions is evident: radio, infra red (IR), optical (near-IR, visible, near-UV), ultra violet (UV), x-ray and gamma. From IR to UV, different sets of \textit{imaging filters} are  used to further subdivide a given range of the electromagnetic spectrum allowing for ``color photography'' of the sky. The goal of imaging sky surveys is to take snapshots of the sky to be able to identify all celestial objects\footnote{Throughout the paper we will refer to discrete physical celestial objects emitting light as \textit{objects} or \textit{sources}. Individual observations of objects (using different filters/instruments or just different epochs) are called \textit{detections}.} in each imaging filter to a given faintness limit.

With the advance of detector technology and growth of the mirror area of telescopes, an exponentially growing area of the sky can be surveyed in a given amount of time. This not only helps map a larger portion of the Universe, but also to take measurements in the time domain.

Multi-wavelength astronomy combines information from different instruments to investigate the physical properties and to constrain theoretical models of celestial objects. Time domain astronomy is the emerging field of the 21$^{th}$ century. While objects of variable brightness have been systematically observed before, the new surveys, like PanSTARRS and LSST, will provide and unprecedented opportunity to find the faint, fast-moving celestial bodies of the Solar System, the serendipitous variable stars of the Milky Way and nearby galaxies. They will help constrain models of quasars and to detect distant supernovae to test our understanding of the dynamics of the entire Universe. 

\subsection{Astronomical catalogs}

Images taken during astronomical surveys are \textit{reduced} by the so called \textit{photometric} or \textit{imaging} pipe-line software. During the reduction process, individual objects are identified in the images and their readily measurable properties are determined, such as integrated brightness, brightness profiles, morphological parameters etc. The number of measured properties is typically in the hundreds.

Because of the distortions introduced by the optical systems of the telescopes and (in case of ground-based telescopes) the turbulent motions in the atmosphere, sophisticated \textit{astrometric} algorithms using non-linear models are needed to precisely determine the coordinates of the detected sources. The quality of astrometry is limited by the goodness of the model used to calibrate positions to well knows standards, and the atmospheric conditions (the so called \textit{seeing}). \textit{Astrometric errors} are usually given in arc seconds (as). The typical error of optical surveys is in the $0.1$-$1.0$~as range. On the other hand, as the direction of gamma photons is much harder to determine, the astrometric error of high-energy transient events, like gamma-ray bursts can be as high as tens of degrees.

Photometric and morphological properties and the astrometric parameters of the identified objects are organized into \textit{catalogs}. Catalogs are traditionally created as plain files listing all objects found in an imaging frame. Since the millennium,  relational database management systems have been widely used to organize, process and mine the information gathered by sky surveys. The numbers of objects detected by the surveys are all on scales, currently topping in the hundred million range. Ongoing and future surveys will provide information about billions of objects, about a hundred detections of each at different times. The typical data volume of current reduced catalogs tops in the $10$~TB range, quickly moving toward hundreds of terabytes, reaching the petabyte range by the end of the decade. The amount of raw imaging data collected and processed during the surveys can be about ten to a hundred times more.

%In this paper, we describe a system that deals with astronomical catalogs loaded into database servers. Although we have complete control over those databases that reside on our system, users can upload their own data, and future versions of the system will be able to incorporate remote data sources into the computations.

When loaded into relational databases, in the simplest case, object detection catalogs occupy one large table, while additional, much smaller tables are used to store metadata. A typical catalog table has an integer primary key, usually composed of different identifiers of the observation. Time and location on the celestial sphere are usually converted to a standardized coordinate system, so conversion between systems is not necessary at runtime. The directions of the detections are often stored as the \mbox{$x,y,z$} Cartesian coordinates of the unit vectors. The rest of the table columns consist of various identifiers, classification parameters (integers) and measured observational parameters (floating point). The tables often have several hundred columns. The frequency of usage of the columns in user queries, however, varies significantly.

\subsection{SQL for astronomical data mining}
\label{sec:sqlreasons}

Since the spread of relational databases in astronomy, the SQL language has become an every day tool of researchers. Its industry-wide support and its ability to describe complex problems with simple syntax in a declarative manner makes it a good choice of programmatic interface to astronomical data warehouses. Although several of the typical data filtering tasks could be done with web forms or other types of custom user interfaces, SQL gives the ability to \textit{script} the operations. This is absolutely necessary for astronomers dealing with data processing issues, as astronomical data usually have to be reprocessed many times during a research project.

To support analysis of astronomical data, extensive libraries of scientific functions have been developed, which can be accessed directly from SQL via user-defined functions. Functions include cosmological distance calculations \cite{2010PASP..122..976T} and various spherical indexing schemes for fast coordinate and region-based searches \cite{2010PASP..122.1375B}.

Together with user-defined functions, data filtering capabilities and aggregate functions, the SQL language turned out to be a very powerful tool for statistical analysis of astronomical data. Being able to solve all problems using SQL makes it possible to process data without pulling it out from the database server.

\subsection{Catalog footprints}

The region of the sky covered by a survey is called \textit{footprint}. Catalog footprints can be exceptionally complex due to observation strategies and the strange geometries of the instruments. \textit{Masks}, also geometric shapes, similar to footprints in complexity, are usually generated from the images to exclude regions of observations of low quality.

Using bitmaps to represent footprints and masks is not favorable because of the limited resolution, the polar singularities and the complicated tessellation schemes of the sphere. On the other hand, bitmaps make it extremely easy to determine the boolean combinations of footprints and masks, once the same pixelization is used.

Budav\'{a}ri et al. developed a software library to describe regions on the surface of the sphere analytically \cite{2010PASP..122.1375B}. Arbitrarily complex regions are constructed as unions of spherical convexes. Convexes themselves are defined as intersections of circles drawn on the surface of the sphere. The resolution that can be reached using double precision arithmetics is about \textcolor{red}{XX~mas}. The library supports exact boolean operations between regions. Clearly, using an analytic approach can be computationally more expensive than using bitmaps, but the much better resolution and the higher flexibility favor the analytical description. We will use the analytic library in our cross-match solution presented in this paper.

\section{Coordinate-based cross-identification}
\label{sec:xmatchintro}

When we have multiple detections of the same celestial source, the measured coordinates will slightly differ. In order to cross-match the detections of two catalogs, we have to measure the distance between all detection pairs, and only accept those pairs as matches that are closer than a given threshold. %In Sec.~\ref{sec:bayesian}, we will briefly introduce the mathematically stringent Bayesian model of cross-identification and show that a simple cut on the separation angle of the detections is sufficient.

In practice, cross-matching is done by excluding obvious false matches first. Different indexing schemes of the sphere have been invented for relational databases to find matching candidates efficiently \cite{2004ASPC..314..289F,2005ApJ...622..759G}. We will briefly explain the so called \textit{zone algorithm} in Sec.~\ref{sec:zone}, which is the fastest available algorithm for Microsoft SQL Server so far \cite{2007cs........1172G}.

There are three different ways a catalog can be cross-matched with other catalogs. One can require that certain catalogs \textit{must} contain good candidate detections of an object in order to accept a match. Additionally, some other catalogs \textit{may} contain detections and should be taken into account, if possible. These are typically catalogs with brighter detection limits. In the third case, one requires that the catalog \textit{must not} contain any candidate detections that would match with detections of other catalogs. This third case is called \textit{drop-out detection}. While the first case is similar to inner joins, and the second case is to outer joins, there is no simple equivalent of the third case in standard SQL.

Drop-out detection is particularly important in multi-wavelength astronomy because even if a certain object is not detected by an instrument due to its too high detection limits, an upper limit to the brightness of the object can still be given based on the known the sensitivity of the instrument. Looking for missing objects can be used to find, for example, a certain kind of galaxy that is bright in the infra red but too faint in ultra violet to show up in the UV images. To safely detect drop-outs, it is fundamentally important to know whether an object is not in the catalog because its celestial location was not observed by the survey at all, it was missed by the instrument because it was too faint or it was intentionally masked out and excluded from the catalog due to other reasons. To account for this problem, an exact description of the observed areas, the so called \textit{footprints}, and \textit{masks} is necessary.

\subsection{Previous work}

The first automatic cross-identification on-line service was implemented by Budav\'ari et al. as a set of XML SOAP web services \cite{2003ASPC..295...31B}. As it was a prototype built to demonstrate the then new web service technology, not much attention to the performance and scaling properties was paid. Based on the idea, an open, SOAP-web-service-based standard, Open SkyQuery was developed by the National Virtual Observatory to federate geographically distributed data sets \cite{2004ASPC..314..177B}. Both of these versions used the SQL language with some custom functions as the main programming interface. Because Open SkyQuery could not benefit from co-located data sets, and due to performance issues the system was limited to process only 5000 matches in a run.

The Virtual Observatory Alliance standardized the Astronomical Data Query Language (ADQL) intended to be used as the \textit{lingua franca} of astronomical catalogs \cite{2011arXiv1110.0503O}. Though the ADQL language defines many new, astronomy-induced constructs compared to SQL, including spherical region expressions that can be used to circumscribe cross-matching problems, ADQL was not designed with query optimizability in mind.

The new CDS xMatch service implements a high performance cross-match engine that partially uses database technology, and can perform two-way joins only \cite{CDSxMatch}. The current version of the service features a form-based user interface only and no scripting support.

\subsection{Probabilistic cross-identification}
\label{sec:bayesian}

The proper statistical formulation of the problem is based on Bayesian probability theory. Every detection has a measured direction unit vector $x_i$ accompanied by its uncertainty, which is determined in the calibration process. Often the catalogs would assume Gaussian errors and quote a single $\sigma$ value, others estimate the precision per detection, $\sigma_i$. These two together form a likelihood function of the unknown model position $m$ on the sky. In the general spherical case, \mbox{$L_{x_i}\!(m)\!=\!F(x_i;m,w_i)$} where $F()$ is the Fisher distribution
\begin{equation}
F(x_i;m,w_i) = \frac{w_i}{4\pi \sinh w_i}\,\exp \big(w_i m x_i\big).
\end{equation}
This is the simplest analog of the Gaussian on the surface of the unit sphere \cite{1953RSPSA.217..295F}, where $w_i$ is its precision parameter, which is $1/\sigma_i^2$ in the limit of high accuracies.

Given a set of detections with their positions on the celestial sphere \mbox{$D=\{x_i\}$} and the corresponding uncertainties, we compare two competing complementary hypotheses: Are the detections from the same source or not? 
In one case the model assumes a single object at an unknown position $m$ and the likelihood of the hypothesis is a result of an integral over all possible $m$:
\begin{equation}
L_{{\rm{}same}} = \int\!dm\ \pi(m)\,L_D(m) = \int\!dm\ \pi(m)\,\prod_i L_{x_i}(m).
\end{equation}
The function $\pi(m)$ is the prior, which can be assumed to be isotropic in most cases.

The complementary hypothesis allows any of the detections to come from a separate source, hence its parametrization consists of a set of directions $\{ m_i \}$, one for each observation. The integral over the entire parameter space falls apart into the product:
\begin{equation}
L_{{\rm{}not}} = \prod_i \int\!dm_i\ \pi(m_i)\,L_{x_i}(m_i).
\end{equation}
The Bayes factor is the ratio of these two likelihoods, $B$=$L_{{\rm{}same}}/L_{{\rm{}not}}$. When this ratio is around unity the data is indecisive but when it is much larger than 1, the data favors the match. Alternatively, when $B$ is close to 0, the evidence points toward separate sources. For the Fisher distribution, the calculation can be done analytically and yields
\begin{equation}
B = \frac{\sinh w}{w} \prod_i \frac{w_i}{\sinh w_i},
\end{equation}
where $w=\big|\sum w_i x_i \big|$. In the limit of high accuracies, the result takes a familiar Gaussian form.

We note that based on the Bayes factor and the density of detections in the given catalogs, it is also possible to define a posterior probability for each match. For more details and the full mathematical discussion, we refer the interested reader to \cite{2008ApJ...679..301B}. In our cross-match engine, we use the above Bayes factor as a discriminator between positive and negative matches.

\section{SQL language extensions for SkyQuery}
\label{sec:sqlext}

Because of the reasons detailed in Sec.~\ref{sec:sqlreasons}, we decided to base our SkyQuery language on SQL and extend the language to support easy expression of cross-identification problems and spatial filtering. Building on the basis of SQL not only makes it easy to learn the extended syntax, but also allows for backward compatibility with traditional SQL queries.

When extending the original SQL syntax, we wanted to avoid any interference with the existing behavior of SQL clauses. This is why we introduced the new \texttt{XMATCH} clause (see Sec.~\ref{sec:xmatchsql}) instead of incorporating its functionality into the standard \texttt{FROM} clause and \texttt{JOIN} operators.

There are some ideas that are worth considering when creating extensions to a declarative query language. Our language extensions were designed such a way that all queries that can be described using the language will be executable and can be optimized efficiently. This is in strong contrast, for example, with the way most GIS systems implement spatial constraints (complex boolean expression in the \texttt{WHERE} clause) where efficient optimization is an issue; or in contrast with the rather flexible ADQL language where even query executability is a problem. While flexible syntax might broaden the range of applications a language can be used for, clever syntax restrictions to certain expressions can ensure that all queries can be optimized easily and executed efficiently, without the cost of losing flexibility. One such syntax restriction is to move certain filtering criteria (for example spatial constraints) from the \texttt{WHERE} clause to a new clause that does not allow for complex logical expressions. Simple implementation is always a main objective, especially in case of scientific projects with limited budgets.

\subsection{Cross-match queries in the relational model}

Cross-identification queries are written to join two are more tables containing detections of celestial objects. The join among these tables is made based on the coordinates of the detections in a probabilistic manner. Tables taking part in the cross-match join must contain columns for the measured coordinates of the detections as the two spherical coordinates or as the three components of Cartesian unit vectors; conversion between the two representations is done automatically\footnote{Theoretically, coordinates and coordinate errors can also be specified as expressions allowing for conversions between different coordinate systems.}. These tables must contain a primary key (composites keys are supported) and may contain the astrometric error and any number of other columns. Also, when a catalog table contains multiple detections for each object, a two way \textit{self-join} can be used to find detections of the same object. Our solution also supports joining in additional tables in the same cross-match queries using traditional joins based on foreign keys. There are no restrictions on additional tables joined in using traditional joins, even sub-queries and table valued functions are supported.

In Sec.~\ref{sec:xmatchintro}, we explained the three ways a catalog table can be matched to other catalogs. These three methods are attributes to the tables themselves and not to the operators joining them. This has to be taken into account when designing the SQL language extensions.

The result of a cross-match query, just like any other SQL query, is a table. Cross-match queries may return any combination of the columns of the joined tables, including columns resulting from expressions. The future versions of the cross-match engine might support aggregate functions as well. Aggregations is particularly interesting as cross-match queries are partitioned and executed on many machines in parallel.

As a result of the cross-identification, the best estimate coordinates and Bayes factors characterizing the goodness of the matches are calculated. Because these values have to be referenced in the select list of the queries somehow, we will introduce a \textit{virtual table} which contains the aforementioned parameters for every N-way match. From the aspect of traditional SQL, one can consider this table as a result set of a table valued function.

As catalog tables might contain a large number of columns that are out of the interest of most users, for space saving reasons, it would be more convenient to vertically split the tables and mirror only the frequently used columns to all cluster nodes. The less frequently accessed parts of the tables could be kept on larger, possibly slower, network accessible storage. We plan to address this problem in a future version of the cross-match engine using a ``lazy join'' algorithm.

\subsection{Zone algorithm for efficient cross-matching}
\label{sec:zone}

The cross-match algorithm we use was developed by Gray et al. and is implemented entirely in SQL to benefit from the query optimizer \cite{2007cs........1172G}. The algorithm partitions the sphere along equally spaced latitude circles into \textit{zones}. The actual matching consist of multiple steps, each producing a new table which is materialized in a staging database. First, a \textit{zone table} is generated for each catalog table which contains a special hash of the coordinates (calculated based on the zones). Clever hashing makes finding candidate matches easier, i.e. finding detection pairs that are close enough to each other. From two zone tables, we generate a \textit{pair table} containing the matching detection candidates. This is the step when we impose the strict angular separation cuts. Then, joining the pair table with the two catalog tables, a \textit{match table} is created. Match tables contain the updated coordinate estimates, parameters to calculate the Bayes factor and all the columns necessary to evaluate the final query.

If more than two tables have to be cross-matched, matching is done iteratively, two catalogs at a time. Catalogs are either processed sequentially, matching the next catalog with the match table of the previous iteration, or done in a cascading way, matching two times two catalogs first (in parallel), then matching the resulting two match tables with each other, and so on. While our current implementation matches catalogs sequentially, future versions might favor the cascading approach to further parallelize processing.

In the future, the cascading approach also can be useful in cases when geographically distributed large data sets have to be matched. One can easily imagine a scenario where certain catalogs are only available at one site while other catalogs are only at another data processing facility. Matching the locally available catalogs first at each site and transferring only the matched results from one site to another would help reducing network traffic significantly. We call this approach \textit{co-location-aware optimization}.

\subsection{Defining the N-way probabilistic join}
\label{sec:xmatchsql}

In Sec.~\ref{sec:sqlext}, we explained why we decided to keep much of the traditional SQL syntax intact and introduce only new clauses. To explain the behavior of the new clauses, we consider Query~\ref{que:sample}. The first half of the query (above the customized \texttt{XMATCH} clause) is in traditional SQL. Data sets listed in the \texttt{FROM} clause are called SDSS, TwoMASS and GALEX after three frequently used astronomical catalogs. Table names are separated from data set identifiers by colons. Each of the listed tables contains observations of galaxies. Each table has an integer field \texttt{ObjID} which is the primary key. Spherical coordinates are stored in the \texttt{RA} and \texttt{Dec} columns\footnote{RA stands for \textit{right ascension}, this is the angle measured around the celestial equator; the equivalent of $\phi$ in traditional spherical coordinates. Dec stands for declination, the angle measured from the equator toward the poles; the equivalent of $\theta$.}. Cartesian coordinates are named \texttt{Cx}, \texttt{Cy} and \texttt{Cz}. Columns denoted with \texttt{mag\_}\textit{x} are brightness measurements of the objects in different imaging filters.

\begin{query}
\begin{lstlisting}
SELECT x.RA, x.Dec,
       s.ObjID, s.RA, s.Dec, s.mag_g, s.mag_r, s.mag_i,
       g.ObjID, g.RA, g.Dec, g.mag_nuv, g.mag_fuv,
       t.ObjID, t.RA, t.Dec, t.mag_J, t.mag_H, t.mag_K
INTO   MyDB:NewResults
FROM   SDSS:PhotoObjAll AS s
       CROSS JOIN GALEX:PhotoObjAll AS g
       CROSS JOIN TwoMASS:PhotoXSC AS t
WHERE  s.Galaxy = 1
XMATCH BAYESIAN AS x
       MUST s ON POINT(s.Cx, s.Cy, s.Cz), 0.1
       MUST g ON POINT(g.Ra, g.Dec), 0.2
       MAY  t ON POINT(t.Ra, t.Dec), 0.5
       HAVING LIMIT 1e6
REGION CIRCLE J2000 180 0 60
\end{lstlisting}
\caption{A sample cross-match query demonstrating the extended SQL syntax.}
\label{que:sample}
\end{query}

The \texttt{FROM} clause simply produces the Cartesian product of the three catalog tables. In traditional SQL, one would write a \texttt{WHERE} clause which filters the Cartesian product leaving only matching detections. Obviously, tables of high cardinality cannot be matched that way. One solution would be to analyze the \texttt{WHERE} clause describing the cross-match criteria and optimize query execution accordingly. Such expression analysis and optimization algorithm is way too complicated. Instead, we introduce the new \texttt{XMATCH} clause that eliminates the need for complex expression analysis and simplifies optimization a lot.

At this point, we would like to point out that a typical cross-identification query not only returns the IDs of the matching detections, but also various columns (or expressions of them) of the original catalog tables. If only the list of matching IDs was returned, at the end users would need to execute an N-way inner join to retrieve these additional columns. Such an N-way join could cost as much (or more) in I/O terms as the entire cross-identification. Doing this N-way inner join at the same time with cross-identification, overall I/O costs can be significantly reduced.

In Query~\ref{que:sample}, the \texttt{BAYESIAN} keyword belongs to the \texttt{XMATCH} clause and defines the method of cross-identification. Currently only Bayesian is supported. The \texttt{AS x} alias is used to make the columns calculated by the cross-matching algorithm being able to be referenced by the rest of the query. Note the first line of Query~\ref{que:sample} and the \texttt{x.RA} and \texttt{x.Dec} column references. These columns will contain the best coordinate estimates computed by the Bayesian algorithm. The same virtual table also returns the Bayes factor as \texttt{x.BF} which can be used to calculate posterior probabilities. %, see Sec.~\ref{sec:priors}.

The \texttt{HAVING LIMIT} clause is required and specifies the minimum value of the Bayes factor for a positive match. Again, we decided to use a custom keyword, instead of incorporating this criterion into the \texttt{WHERE} clause to make implementation simpler.

\subsection{Region constraints}
\label{sec:region}

As it was mentioned in Sec.~\ref{sec:xmatchintro}, precise information about the footprint of the catalogs is required to run certain cross-identification queries. Similarly, queries might be restricted by the users to a certain area of the sky. This is demonstrated in the last line of Query~\ref{que:sample}, where the region of interest is restricted to a circle on the sky centered on the coordinate $(180, 0)$ with a radius of $60$~am (arc minute). The \texttt{J2000} prefix defines the equinox of the coordinate system.

An IVOA standard exists \cite{2011arXiv1110.0504R,2010PASP..122.1375B} to describe spherical regions as string, similar to the sample in Query~\ref{que:sample}. These descriptions, however, become increasingly verbose as the regions get more complex. Consequently, a future syntax extension will have to support both inline region descriptions (as in the sample) and a way to reference regions already stored somewhere in some standardized format. One repository of region descriptions could be based on Footprint Services developed earlier by our group \cite{2007ASPC..376..559B}.

According to the considerations we made in Sec.~\ref{sec:sqlext}, language extensions to restrict cross-identification queries to a given area of the sky should use a special construct, most favorably a new clause to describe the spatial filtering criteria. Mixing it with the \texttt{WHERE} clause conditions, where boolean algebraic combinations of various types of filtering criteria are allowed, would make query optimization significantly more complicated. For example, a spatial constraint could be combined with another constraint using the \texttt{OR} operator which would make it very hard to determine at which point of the query execution the spatial constraint has to be applied. On the other hand, if the spatial constraint is not combined with any other criterion, it can be imposed in the very first step of the zone-based algorithm, when the zone table of the first table is being built for the very first catalog joined in in the query, see Sec.~\ref{sec:zone}.

\subsection{Cuts on posterior probability}
\label{sec:priors}

The previously introduced syntax allows for limiting matches based on the Bayes factor. The latter is calculated solely on the basis of measured coordinates and astrometric errors. It might be necessary, however, to further filter matches based on parameters of the detections other than the coordinates. For instance, stars and quasars may look very similar in images but their colors are very different. If a star and a quasar appear very close to each other, the measured coordinates of their detections alone might not be enough to associate the detections with the real objects, so we would end up with false positive matches mixing detections of the star with those of the quasar. Incorporating colors into the model, however, allows for easy separation. In the Bayesian framework introduced in Sec.~\ref{sec:bayesian} probabilistic models can be very easily extended. In the current version, we only support cuts on the Bayes factor, but additional cuts on posterior probabilities can be easily imposed in the \texttt{WHERE} clause of the queries.

\section{Hardware and software setup for SkyQuery}
\label{sec:hwsw}

\subsection{Hardware platform}

Our current system consists of five Dell Power Edge 2950 servers with two Xeon E5430 processors and $24$~GB of RAM each. The I/O subsystem consists of two Dell PERC 5/e RAID controllers connected to two Dell PowerVauld MD1000 storage units containing 15 disk drives each. Disk drives are configured in four 7-disk RAID~5 volumes. RAID~5 is not an evident choice for a high performance system. According to our tests, however, the sequential read bandwidth of the RAID~5 volumes saturates at the same value as the speed of simple striped volumes. Write speed is about half of that, but since our workload is highly biased toward sequential reads, this did not happen to have a huge impact on the overall performance. The benefit from using RAID~5 volumes is the increased directly attached storage capacity of the servers. The total capacity of the I/O system is $16$~TB per server with a sustained sequential read throughput of $1.2$~GB/s. The system is supposed to be very easily scaled up to higher performance servers according to the scaling abilities of SQL Server.

The cluster nodes are connected with $10$~GB/s network links using IP protocol. The database servers are supplemented by a head node dedicated to coordinating job execution and running the web server for the user interfaces. The head node also contains a central database storing the state of the cluster.

\subsection{Software platform}

We based the implementation on Microsoft SQL Server 2008 R2, Windows Server 2008 and the \dotnet \ platform. Our decade long experience with the SQL Server line, and many existing libraries written in \csharp \ for the \dotnet \ Framework made these the obvious choice. Porting the system to other platforms would require rewriting a major fraction of the code including legacy libraries. While most of the cross-identification algorithm is implemented in pure SQL, \csharp \ code is used to generate the queries on the fly. We extensively use user-defined functions running inside the SQL CLR to perform region based calculations, see Sec.~\ref{sec:region}. To implement the parallel job execution and queueing system, see Sec.~\ref{sec:jobs}, we rely on version~4 of \dotnet \ Workflow Foundation. Its massive support for parallel activities, automatic workflow persistence and integration with the \dotnet \ Framework made it the best candidate to implement our query pipelines. 
%If not otherwise cited, all other modules and libraries used for the system are the work of the authors of the paper.

\subsection{Database setup}
\label{sec:dblayout}

Database layouts are optimized for the underlying hardware. Database tables of the astronomical catalogs are organized into file groups (separate file group for each large table) to optimize for long table scans that happen when cross-matching entire catalogs. File groups contain multiple files split across the RAID volumes. The database server makes sure that data is evenly distributed among the database files.

Databases storing catalog data are mirrored to every cluster node to allow for parallelization and load-balancing. These databases are all set to read-only. We create a so called \textit{mini} version of every catalog to support gathering query statistics on the fly. Mini databases are uniformly sampled from the original databases at a $10^{-3}$ sampling rate. Sampling is done such a way that foreign key references remain intact. In the case of astronomical catalogs it usually can be achieved by sampling the object table first and simply enforcing foreign key constraints on the rest of the tables.

SkyQuery uses staging databases to store intermediate output produced by the cross-identification algorithm, see Sec.~\ref{sec:zone}. These databases are heavily used for both reading and writing. Since in SQL Server transaction logging cannot be turned off completely, we optimize all queries writing their results into the staging databases to do minimal logging. Currently we have one staging database per cluster node, but the API supports automatic selection of high speed background storage volumes, like SSD-based arrays for staging, if necessary.

As users interact with the system via SQL, the most convenient way to store query results is to allocate a moderately sized database for each user, and save query results there \cite{2004ASPC..314..372O}. Users can also upload their own data tables and store them in their so called \textit{MyDB}. The final result sets can be easily downloaded as files. In the current configuration, MyDBs are distributed among the cluster nodes; only one copy of a database per user. In addition to table storage, in the final system, users will be able to create their own views and write their own user-defined functions as well. 

A future version will give the users restricted access to the big staging databases on the cluster nodes. This is important in cases when the results of a computation require only limited storage but internal steps might produce larger outputs.

\section{Implementation details}
\label{sec:implementation}

\subsection{Cluster Management}

Unfortunately, Microsoft SQL Server does not support clustering of database servers other than fail-over clustering. Although there exists support in SQL Server for linking servers over a network connection and execute cross-server queries, the performance is poor. Consequently, we have to rely on single server databases and build our solutions around this limitation. We target two types of problems: a) partition databases too big to fit on a single server and distribute them over a set of servers, and b) mirror exact copies of databases to a set of servers to distribute the load of small queries and to parallelize big queries.

To be able to easily develop advanced federated database solutions, first we built a \textit{cluster management} application and API that can register and manage all the necessary information about the hardware and software configuration, database layouts, and users of the cluster. The registry stores detailed information about the individual server machines. Machines are organized by roles, so different hardware configurations can be assigned to different tasks. Information about the I/O system is also stored, including the size, bandwidth, fail-over level etc. of the logical disk volumes. Disk volumes can be easily assigned to various tasks like storage, staging, temporary, log etc. This becomes increasingly important with applications having both CPU-bound and I/O-bound components that should be targeted to the most appropriate hardware.

We organize databases into \textit{federations}. This is a loosely bound set of databases belonging to a certain application. Federations can contain \textit{database definitions}. A database definition is not an actual database but a prototype with all the schema but no data. Database definitions also contain information about how the data should be distributed over the cluster machines. Actual physical databases are created based on these prototype schemas automatically by the API. 
%\textcolor{red}{How file layout is determined.}

The cluster management API also contains functionality to manage users associated with the federations and a job queueing system we will describe in Sec.~\ref{sec:jobs} in detail.

\subsection{Query parsing and identifier resolution}

To parse the special extensions introduced to the SQL language in Sec.~\ref{sec:sqlext}, we wrote a parser generator from scratch that we use to generate a parser from the extended grammar described in BNF. The main reason behind writing our own solution instead of using common parser generators was that we wanted to use certain features of the \csharp language that was not supported by other generators.

Once a query expression is parsed and the parsing tree is built, identifiers referencing tables and columns are resolved based on the underlying database schema. At this point we can collect all the information needed to execute a query. First, tables and databases referenced by the query are identified. Although all large databases are mirrored to the cluster nodes, cross-identification queries may reference tables from the MyDB of the user which is only available on one of the nodes. All tables in MyDB are assumed to be small enough so they can be simply cached in the staging databases of the worker nodes prior to query execution. Future versions of the system will have the ability to fetch data from remote, Virtual Observatory compatible data sources a similar way, via the Internet.

Before proceeding to the query optimization step, sanity checks are performed to make sure all tables referenced in the query conform with the requirements of the cross-match algorithm, for example all tables have appropriate primary keys and the data types of the columns storing spherical coordinates are compatible.

\subsection{Query optimization and partitioning}

The zone algorithm cross-matches catalogs pairwise. Once two catalogs have been cross-identified, the new best estimate coordinates are calculated and passed on to the next iteration. From the perspective of optimization, starting with catalogs of the least cardinality is the best choice, except when looking for drop-outs, see Sec.~\ref{sec:xmatchintro}~and~\ref{sec:zone}.

Since the extended SQL syntax supports filtering the data, and spatial constraints can also be applied to the queries, there is not much use to store static cardinality information about the catalogs. Instead, our system is designed to gather statistics about each query prior to optimization. In Sec.~\ref{sec:dblayout}, we mentioned that random subsets of the source catalogs, the so called \textit{mini} databases are created and stored on the cluster nodes. We use these mini catalogs to get quick statistics about the source tables referenced by the queries. Once the tables are identified in the parsing tree, all criteria restricting the rows of that table are also collected from the \texttt{ON} conditions of the \texttt{JOIN} expressions and from the \texttt{WHERE} clause of the query. From this information we are not only able to estimate the cardinality of the source tables but also to determine the spatial distribution of the object detections of the astronomical catalogs after all the selection criteria have been applied.

Information about the spatial distribution of the data points is essential in order to be able to efficiently partition the query. As the zone algorithm indexes the surface of the sphere based on the declination (latitude) angle, partitioning on the right ascension (longitude) angle is a convenient choice. Based on the histogram of right ascensions, the surface of the sphere is split into disjoint partitions defined by great circles intersecting at the poles. Partition boundaries can be chosen anywhere as all the data is mirrored to every cluster node. This also eliminates the need of buffer zones along partition boundaries. Boundaries are chosen such a way that an approximately equal number of detections fall into each partition.

The number of partitions is chosen to be a multiple of the available cluster nodes. We use more partitions than the number of available physical machines to execute the cross-match task, because higher granularity makes error recovery much easier, only smaller parts of the job had to be redone when unexpected events happen.

\subsection{Jobs as parallel workflows}
\label{sec:jobs}

Every partition of a cross-match query translates to a sequence of ordinary SQL queries, and these query sequences run in parallel on many machines. Because of the complexities of multi-threaded application development, we decided to implement the cross-match jobs as \textit{workflows} written for \dotnet \  Workflow Foundation (WF) version 4. WF has extensive support for parallel execution of \textit{activities} (atomic components of workflows), and also for exception handling and workflow cancellation logic.

Workflows make it sure that ordinary SQL queries performing the cross-matching will run in the necessary order and that partitions will be processed in parallel. All ordinary SQL queries are written such a way that they do not return any data but write all results into the staging databases of the particular cluster nodes. Also, all heavy computations are coded into these SQL queries. These design constraints make it possible to build workflows that do only basic computations and issue regular SQL queries to remote servers to do the rest. As a result, all workflows can be run on a single head node of the cluster instead of scheduling non-SQL user code on the cluster nodes.

For each cross-match query, we create a new job (in the form of a WF workflow) and schedule its execution with a custom-written queueing system. The queueing system supports execution of jobs with different time-out intervals. This is particularly important in open access database systems, like SkyQuery, where queries written by the users can have any complexity. Users developing queries would submit them first to the \textit{quick} queue to see if the queries work correctly on smaller chunks of the data. Once they are satisfied with the results, they can send the queries to the long queue with much longer time-out interval for guaranteed completion.

To avoid moving large amounts of data between servers, we schedule the execution of an entire partition of a cross-identification job on the same cluster node. Cluster nodes are assigned to the partitions in a round robin fashion. We chose round robin scheduling over complex load balancing because of some problems arising from the behavior of database servers. For instance, it is hard to correctly measure the load on a particular database server as either CPU load or I/O load can vary heavily during query processing. Assigning servers to tasks in round robin seems a much simpler and reasonable way.

We extended the WF base library with a custom activity to support re-execution of a certain branch of a workflow upon an exception. The typical scenario we wanted to cover is when a single cluster node goes down and a regular SQL query fails. In this case, another cluster node can be assigned to the failing branch of the workflow and the branch can be re-executed without affecting other branches of the entire workflow. The number of re-tries can be limited, so permanent errors in the system won't cause the workflows to go into an infinite loop but to fail permanently.

Because we expect some really long running jobs, we had to deal with the issue of suspending and resuming jobs for system maintenance reasons. Although WF supports persisting the state of workflows, a workflow can only be suspended at transition points between activities. When a suspend request arises, most workflows can be suspended in a short time interval. Some activities executing longer-running ordinary SQL queries that do not complete within the time-out period are simply cancelled and restarted whenever the system comes on-line again. This is another reason to use partitions of higher granularity. Since ordinary SQL queries cannot be suspended, only cancelled, partitioning is a good way to limit the size of operations done in a single step, thus to shorten single query execution times.

To maintain the integrity of the system and to save temporary and staging storage space, all job workflows are designed to fail gracefully and to be able to be cancelled gracefully, i.e. they remove all temporary data generated and restore the original state of the system.

Job workflows are implemented and installed as \dotnet \ binary assemblies, so any change to the workflows requires recompilation of the assemblies with higher version numbers. We designed the job queueing system to be able to handle multiple version of workflow assemblies, so changes to the workflows or activities will not require a system restart.

\subsection{Bulk data operations}
\label{sec:bulkup}

Since we have to deal with large amounts of data we had to optimize all data moving operations, especially those that happen between machines. We implemented a service that runs on all servers of the cluster and handles bulk data copy requests from the head node. Entire data tables or subsets of tables are copied across servers using SQL Server bulk-copy, while entire databases are copied using robust file copy.

We implemented large copy jobs as workflows. For instance, a workflow was written to mirror databases to all cluster nodes in a cascading manner: first a single copy of the original database is made, then the two copies are used to make two more mirrors in parallel, and so on.

\subsection{Metadata management}

The relational data model itself only uses table names and column names to identify quantities stored in the databases. Scientific applications, on the other hand, require detailed description of the physical quantities. The Internatinal Virtual Observatory Alliance defines the ontology and metadata models to describe astronomical data. In SQL Server, extended properties can be added to every database schema object. These properties can be easily queried via special views. For SkyQuery, we use these extended properties to store meta-information. Metadata includes description of the quantities using identifiers based on the ontology but also human-readable text to display on web pages, etc.

Since all data in our system are manipulated with SQL scripts, results of the computations are manifested as output tables stored in the users' MyDBs. Because we parse every SQL query executed we have complete control over the schema and metadata of the output tables as well. It will be very convenient in the future to derive metadata and provenance information about query outputs directly from the SQL scripts.

%\subsection{User interface}

%\textcolor{red}{Erre nincsen hely, nem annyira fontos, hagyjuk ki.}

\section{Summary and future work}
\label{sec:summary}

In this paper we have introduced a new, scalable implementation of software for cross-identification of co-located astronomical catalogs. Compared to the earlier reincarnations, for the third version of SkyQuery, the following improvements have been made. a) Instead of single-server operation, queries are partitioned and executed on a cluster of identical database servers having identical versions of all data sets. b) An easy to optimize syntax extension to the SQL language was invented to support simple formulation of cross-match problems. c) Queries are translated into complex workflows of traditional SQL queries. Workflows are implemented in Windows Workflow Foundation to support parallel execution.

For the next versions of the system, we are working on the following additions. a) Right now, all queries are run from scratch, i.e. helper tabled used to speed up cross-matching are newly created every time when needed.  Certain helper tables, e.g. zone tables, could be cached to further speed up execution. b) We are designing a generic framework for handling metadata which will allow extracting provenance information directly from the queries written by users. c) A lazy-join algorithm is being designed to allow vertically partition tables. This will make it possible to move less frequently used columns to cheaper storage. d) We will add support to reference tables from remote data sets accessible to Virtual Observatory standard protocols. In the future, we plan to update the system to be able to cooperate with remote data centers and support co-location-aware query optimization.

\section*{Acknowledgements}

This work was supported by the following Hungarian grants: NKTH: Pol\'anyi and KCKHA005.

The Project is supported by the European Union and co-financed by the European Social Fund (grant agreement no. T\'AMOP 4.2.1./B-09/1/KMR-2010-0003)



\begin{thebibliography}{1}
\providecommand{\url}[1]{\texttt{#1}}
\providecommand{\urlprefix}{URL }

\bibitem[BPD1]{CDSxMatch} T.Boch, Pineau, F.X., Derriere, S.: {CDS} {xMatch} service documentation (2011)

\bibitem[Bud1]{2003ASPC..295...31B} Budav{\'a}ri, T., 
Malik, T., Szalay, A.~S., Thakar, A.~R., 
\& Gray, J.\ 2003, ADASS XII, 295, 31 

\bibitem[Bud2]{2004ASPC..314..177B} Budav{\'a}ri, T., Szalay, A.~S., Gray, J., et al.\ 2004, ADASS XIII, 314, 177 

\bibitem[Bud3]{2007ASPC..376..559B} Budav{\'a}ri, T., Dobos, L., Szalay, A.~S., et al.\ 2007, ADASS XVI, 376, 559 

\bibitem[Bud4]{2008ApJ...679..301B} Budav{\'a}ri, T., \& Szalay, A.~S.\ 2008, \apj, 679, 301 

\bibitem[Bud5]{2010PASP..122.1375B} Budav{\'a}ri, T., Szalay, A.~S., \& Fekete, G.\ 2010, \pasp, 122, 1375 

\bibitem[Fek]{2004ASPC..314..289F} Fekete, G., Szalay, A.~S., \& Gray, J.\ 2004, ADASS XIII, 314, 289 
  
\bibitem[Fis]{1953RSPSA.217..295F} Fisher, R.\ 1953, Royal Society 
of London Proceedings Series A, 217, 295

\bibitem[Gor]{2005ApJ...622..759G} G{\'o}rski, K.~M., Hivon, E., Banday, A.~J., et al.\ 2005, \apj, 622, 759 

\bibitem[Gra]{2007cs........1172G} Gray, J., Szalay, A., Budavari, T., et al.\ 2007, arXiv:cs/0701172 

\bibitem[OM1]{2004ASPC..314..372O} O'Mullane, W., Gray, J., Li, N., et al.\ 2004, ADASS XIII, 314, 372 

\bibitem[Or1]{2011arXiv1110.0503O} Ortiz, I., Lusted, J., Dowler, P., et al.\ 2011, arXiv:1110.0503 

\bibitem[Ro1]{2011arXiv1110.0504R} Rots, A.~H.\ 2011, 
arXiv:1110.0504 

\bibitem[TP1]{2010PASP..122..976T} Taghizadeh-Popp, M.\ 
2010, \pasp, 122, 976 

\end{thebibliography}
\end{document}